\documentstyle[12pt,epsfig]{article}
\newcommand{\be}[1]{\begin{equation} \label{(#1)}}
\newcommand{\ee}{\end{equation}}
\newcommand{\ba}[1]{\begin{eqnarray} \label{(#1)}}
\newcommand{\ea}{\end{eqnarray}}

\def\pmb#1{\setbox0=\hbox{#1}  \kern-.015em\copy0\kern-\wd0
  \kern.03em\copy0\kern-\wd0
  \kern-.015em\raise.0233em\box0 }

\def\rp{$R_p\hspace{-1em}/\ \ $}

\def\lg{\langle}
\def\rg{\rangle}

\begin{document}
\begin{titlepage}

\begin{center}
{\large\bf Lepton Flavor Violation: Constraints from exotic
muon to electron conversion}
\footnote{Based on the Invited talk given by T.S. Kosmas at
the {\it International Conference on Non-Accelerator New
Physics(NANP'99)}, Dubna, Russia, 1999.}\\

\bigskip

T.S. Kosmas $^a$ and
Sergey Kovalenko $^{b,}$\footnote{On leave from the Joint Institut for
Nuclear Research, Dubna, Russia. }
\\[0.5cm]

{$^a$\it Division of Theoretical Physics, University of Ioannina GR-45110
Ioannina, Greece}\\[0.3cm]
{$^b$\it Departamento de F\'\i sica, Universidad
T\'ecnica Federico Santa Mar\'\i a, Casilla 110-V, Valpara\'\i so, Chile}

\end{center}

\bigskip

\begin{abstract}
The exotic neutrinoless $\mu^-- e^-$ nuclear conversion is studied
within the conventional extensions of the standard model as well as
in the minimal supersymmetric (SUSY) models with and without
R-parity conservation. The dependence of the $\mu -e$
conversion rates on the nucleon and nuclear structure is consistently
taken into account. Using our calculated transition matrix elements
and the available experimental data on the branching ratio $R_{\mu e^-}$
for $^{48}$Ti and $^{208}$Pb as well as the expected experimental
sensitivity for $^{27}$Al employed as a target in the planned at
Brookhaven $\mu^--e^-$ conversion (MECO) experiment, we
extract very severe constraints for the flavor violation parameters.
We especially emphasize on the constraints resulting for SUSY R-parity
violating parameters.
\end{abstract}

\end{titlepage}
\bigskip
\section{Introduction}

The lepton flavor violating (LFV) neutrinoless conversion of a bound 1s-muon
to electron in the field of a nucleus \cite{Marci}-\cite{kos94}
        \begin{equation}\label{reaction}
        (A,Z) + \mu^-_b \to e^- + (A,Z)^*,
        \end{equation}
is known as one of the best probes to search for the hypothetical muon and
electron flavor non-conservation \cite{Marci}. So far, the experiments
\cite{Dohmen}-\cite{Molzon}
seeking for $\mu-e$ conversion events have only succeeded to put upper bounds on
the branching ratio
\begin{eqnarray}
\label{def1}
R_{\mu e^-}=\Gamma(\mu^-\to e^-)/\Gamma(\mu^-\to\nu_\mu),
\end{eqnarray}
i.e. the ratio of the muon-electron conversion rate
relative to the total rate of the ordinary muon capture.
The best upper limits have been extracted at PSI
by the SINDRUM II experiments in the values
(at 90\% confidence level)
\begin{equation}
\label{Rme-Ti}
R_{\mu e^-} < 7.0\times 10^{-13} \quad\mbox{ for }^{48}\mbox{Ti
        target \cite{Dohmen}},
\end{equation}
\begin{equation}
\label{Rme-Pb}
R_{\mu e^-} < 4.6\times 10^{-11} \quad\mbox{ for }^{208}\mbox{Pb
        target \cite{Honec}}.
\end{equation}
These limits are improvements over the previous limits set
at TRIUMF \cite{Ahmad} a decade ago using the same targets.

At present, two $\mu -e$ conversion experiments are launched,
the ongoing experiment at PSI using $^{48}$Ti target \cite{Dohmen},
and the planned MECO experiment at Brookhaven \cite{Molzon}
using $^{27}$Al target. The expected sensitivity on $R_{\mu e^-}$ in
the PSI experiment is $10^{-14}$ while in the Brookhaven
experiment it will be roughly
\begin{equation}
\label{Rme-Al}
R_{\mu e^-} < 2\times 10^{-17} \quad \mbox{for}^{27}\mbox{Al
        target \cite{Molzon}},
\end{equation}
which implies an improvement over the existing limits by about four orders
of magnitude.

The MECO experiment is going to be conducted in a new $\mu$ beam
line at the AGS, where the muons are produced using a pulsed
proton beam \cite{Molz99}. The proton energy
will be chosen in the range of 8-20 GeV to optimize the $\mu^-$ flux
per unit time. Furthermore, the number of electrons with energy
$E_e = 104$ MeV, equal to the energy for the coherent peak in
$^{27}$Al, is very much suppressed. This
is in contrast with the case of $\mu^- \to e^- \gamma$, where the electron
flux from $\mu^- \to e^- \nu {\bar \nu}$ decay is peaked at the energy of
the electrons from $\mu^- \to e^-\gamma$.

For such experiments the knowledge of nuclear transition matrix elements
for all accessible $\mu^-\to e^-$ channels of the targets employed are of
significant importance \cite{KV90}-\cite{SKF98}.
In this work we use the transition matrix
elements calculated for the aforementioned isotopes in the coherent mode
to constrain the lepton flavor violating parameters of various Lagrangians
predicting this exotic process (e.g.\ scalar and vector couplings,
neutrino mixing angles and masses \cite{KFV97}-\cite{R-cons},
supersymmetric R-parity violating couplings
\cite{Kim-Ko}-\cite{Kos-Kov} etc.).
To this aim the recent experimental data and
the expected sensitivity of the MECO experiment.

As is well known, only the coherent rate can be measured because it is free
from background events from bound muon decay and radiative muon capture
followed by a fully asymmetric
$e^+ e^-$ pair creation \cite{Molzon}. On the other hand,
previous studies of $\mu -e$ conversion rates \cite{KVF98}-\cite{Chiang}
have shown that for all mechanisms the coherent mode dominates the process
(\ref{reaction}) which means that this is the most important channel. The
incoherent reaction leading to excited nuclear states is suppressed
due to Pauli blocking effects and it is much harder to calculate
but its knowledge is also useful in order to determine the experimentally
interesting quantity of the ratio of the coherent to the total
$\mu^--e^-$ rate (see Refs. \cite{KV90}-\cite{SKF98}).

\section{$\mu^--e^-$  conversion within common extensions of
Standard Model }

The family quantum numbers $L_e$, $L_{\mu}$, $L_\tau$ are conserved within
the standard model
(SM) in all orders of perturbation theory. However this is an accidental
consequence of the SM field content and gauge invariance. Physics beyond the
SM can easily spoil this property.
Processes like $\mu-e$ conversion, which is forbidden in the standard
model by muon and electron quantum number conservation, play an important
role in the study of flavor changing neutral currents and possible physics
beyond the SM.

On the particle physics side \cite{Marci,JDV86}, there are many
mechanisms of the $\mu-e$ conversion constructed in the literature
(see \cite{Kim-Ko}-\cite{Kos-Kov} and references therein).
All these mechanisms fall into two different categories:
photonic and non-photonic. Mechanisms from different classes significantly
differ from the point of view of the nucleon and nuclear structure
calculations. This stems from the fact that they proceed at different
distances and, therefore, involve different details of the structure.

The long-distance photonic mechanisms are mediated
by the photon exchange between the quark and the $\mu-e$-lepton currents.
These mechanisms resort to the lepton-flavor non-diagonal
electromagnetic vertex which is presumably induced by
the non-standard model physics at the loop-level.
The $\mu-e$ lepton-flavor violating loop can appear
as the $\nu-W$ loop [Fig.1(a)]
with the massive neutrinos $\nu_i$ and the loop with the supersymmetric
particles such as the neutralino(chargino)-slepton(sneutrino) [Fig.1(d)].
In the R-parity violating SUSY models there are also lepton-slepton
and quark-squark loops generated by the superpotential couplings
$\lambda L L E^c$ and $\lambda' L Q D^c$ respectively.

The short-distance non-photonic mechanisms contain heavy
particles in intermediate states and can be realized at
tree level [Fig.2],  at 1-loop-level [Fig.1(a,b,d,e)]
or at the level of box diagrams [Fig.1(c)].
The tree-level diagrams can be constructed in the R-parity
violating SUSY models with the virtual Z-boson,
squarks $\tilde u, \tilde d$ and
sneutrinos $\tilde\nu_i$ [Fig.2].
The 1-loop diagrams of the non-photonic mechanisms include the diagrams
similar to those for the photonic mechanisms but with the Z-boson instead of
the photon [Fig.1(a,d)] as well as additional Z-boson couplings to the
neutrinos and neutralinos [Fig.1(b,e)].
The box diagrams are constructed of the W-bosons and massive neutrinos
[Fig.1(c)] as well as similar boxes with neutralinos and sleptons
or charginos and sneutrinos.

\begin{figure}[h!]
\vspace{-4cm}
\vspace{-1.cm}
\hspace{1.0cm}
\mbox{\epsfxsize=16. cm\epsffile{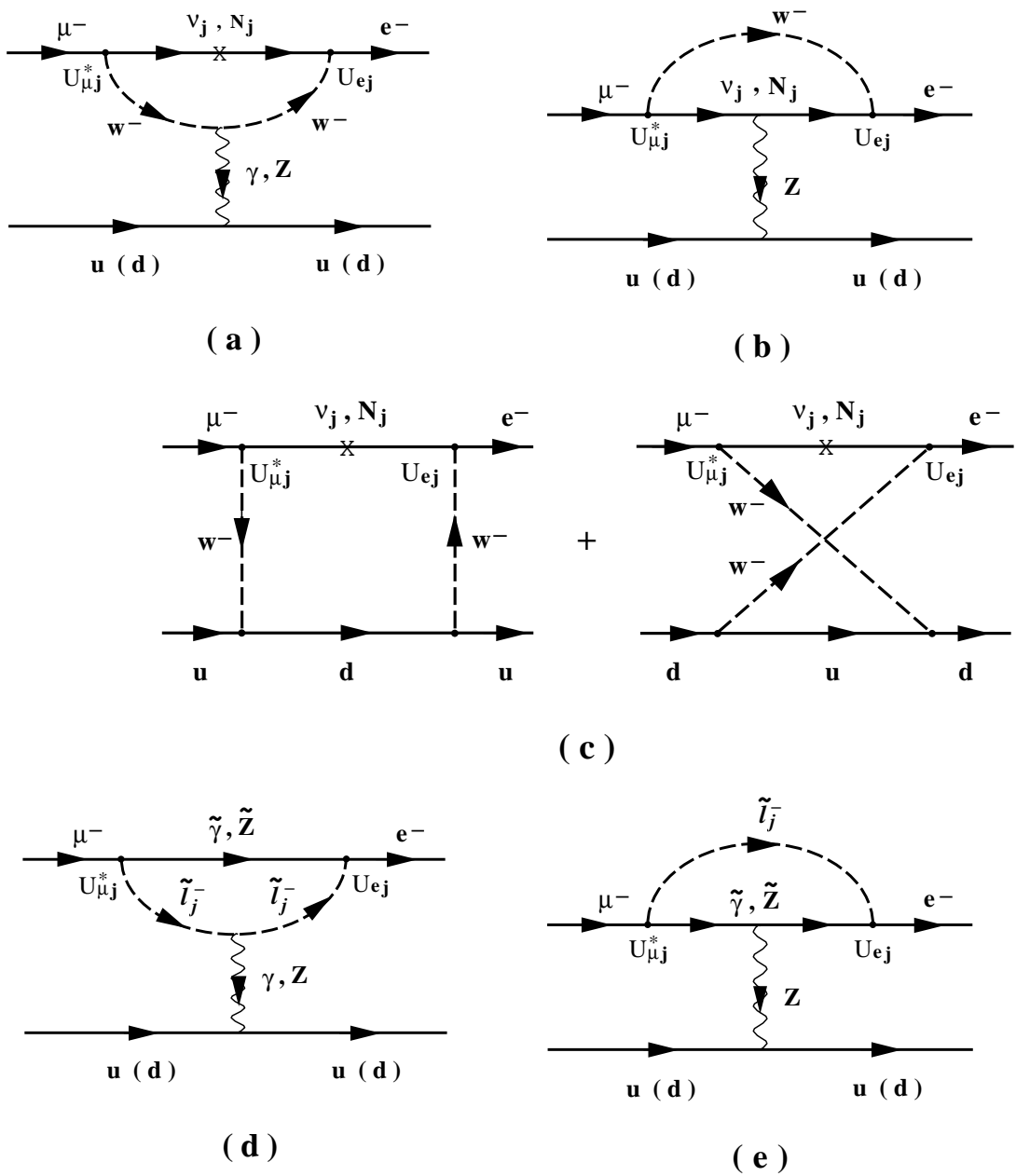}}
\vspace{-3cm}
\caption{
Photonic and non-photonic mechanisms exhibiting the $\mu^-- e^-$
process within the context of conventional extensions of the standard
model (a-c), as well as supersymmetric theories (d,e) with R-parity
conservation. The hadronic vertex is mediated by photon exchange (a,d),
$Z$-particle exchange (a,b,d,e), and $W$-boson exchange (c).}
\end{figure}

Our purpose is to calculate the contribution of
the above-described mechanisms to the  $\mu^--e^-$ conversion
branching ratio (\ref{def1}).
At the first step we construct the effective
$\mu^--e^-$ conversion Hamiltonian for these
mechanisms in terms of nucleon degrees of freedom of a nucleus
involved in the process. This will allow us to accomplish
the calculation of $R_{\mu e^-}$ by applying the conventional nuclear
structure methods using non-relativistic impulse approximation.

\begin{figure}[h!]
\vspace{-5.cm}
\hspace{-2.58cm}
\mbox{\epsfxsize=19. cm\epsffile{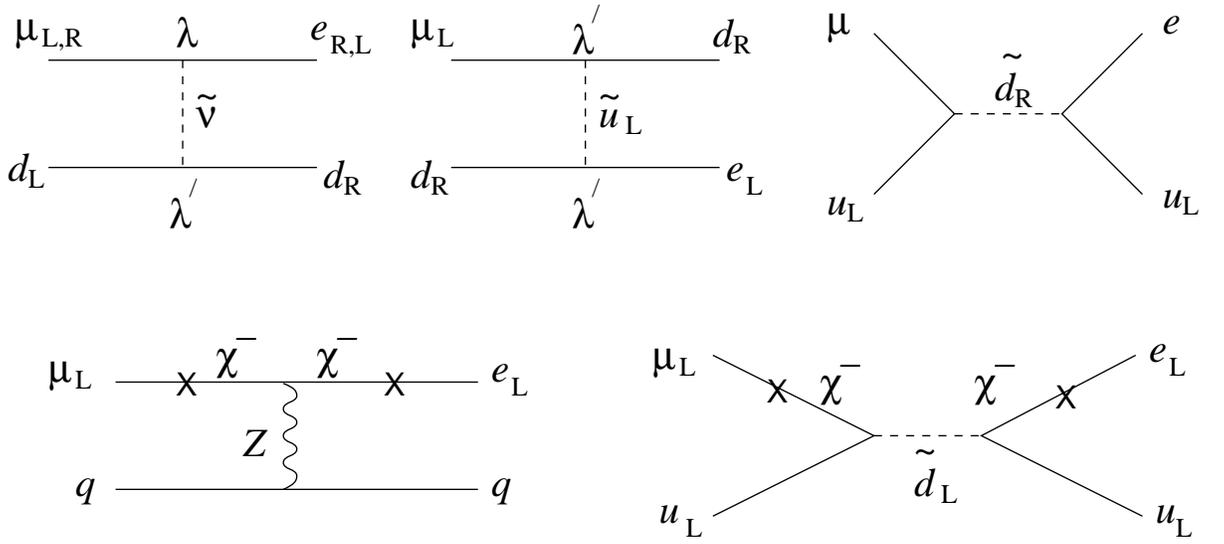}}
\vspace{-14cm}
\caption{
Leading \rp~MSSM diagrams contributing to
$\mu-e$ conversion at tree-level. The chargino-lepton mixing
is schematically denoted by crosses (X) on the lepton lines.}
\end{figure}

For the photonic diagrams of Fig. 1(a,d) the hadronic
vertex represents the usual nucleon electromagnetic current
\begin{equation}
J^{(1)}_{\lambda} = {\bar N}_p \gamma_{\lambda} N_p
 \,= \,{\bar N} \gamma_{\lambda} \frac{1}{2}\,(1 + \tau_3)\, N
\qquad (photonic )
\label{3.1.3}
\end{equation}
$N$ is the nucleon isospin doublet $N^T = (N_p, N_n)$ with $N_p$, $N_n$
being the proton and neutron spinors.

In the case of the non-photonic mechanisms of Fig. 1
the hadronic currents are either the SM neutral currents coupled to
Z-boson [Fig. 1(a,b,d,e)] or the effective currents derived after
an appropriate Fierz transformation from the effective operators of
the box diagrams [Fig. 1(c)] involving heavy particles.
These currents can be written in terms of the nucleon doublet
field as
\begin{equation}
J_{\lambda}^{(2)} = {\bar N}\gamma_{\lambda} \frac{1}{2}\left[
(3 + f_V \beta \tau_3) - (f_V\beta^{''}+f_A \beta^{'} \tau_3)
\gamma_5 \right ] N \qquad (non-photonic)
\label{3.1.4}
\end{equation}
where $f_V$, $f_A$ represent the vector, axial vector static nucleon form
factors ($f_A/f_V = 1.24$) and the parameters $\beta$, $\beta^{\prime}$,
$\beta^{\prime \prime}$, for the models adopted take the values given in
Ref. \cite{KFV97}. The parameter $\beta$ is defined as $\beta =\beta_1/\beta_0$,
with $\beta_0$ $(\beta_1)$ the isoscalar (isovector) couplings at the quark
level. We should note that, in general, the parameters $\beta$, $ \beta^{'}$
and $\beta^{''}$ are functions of $sin^2\theta_W$. For example, in the case
of Z-exchange we have $\beta^{'}=3/2sin^2\theta_W=6.90$.

The corresponding leptonic currents for the photonic and non-photonic mechanism
of Figs. 1 are given in Ref. \cite{KFV97}.

In the minimal SUSY model with a most general form of R-parity violation
(for review see, for instance, \cite{Rp-rev} )
all the possible tree-level diagrams for the $\mu^--e^-$ conversion process
are shown in Fig. 2 \cite{Kim-Ko},\cite{Kos-Kov}.
To write down the low-energy $\mu-e$ conversion Lagrangian
at quark level one starts from the well known R-parity violating
superpotential and integrates out heavy intermediate fields.
For the diagrams of Fig. 2 the corresponding 4-fermion low-energy
effective Lagrangian at the quark level takes the form \cite{Kos-Kov}
(first order of perturbation theory)
\begin{eqnarray}
{\cal L}_{eff}^{q} \,\,\, = && \frac{G_F}{\sqrt{2}}j_{\mu }
\left[ \eta _{L}^{ui}J_{uL(i)}^{\mu }+
\eta_{R}^{ui}J_{uR(i)}^{\mu }+\eta_{L}^{di}J_{dL(i)}^{\mu }+
\eta_{R}^{di}J_{dR(i)}^{\mu}\right]
\nonumber \\
&&+ \frac{G_F}{\sqrt{2}}\left[
\bar\eta_{R}^{di}J_{dR(i)}j_{L} + \bar\eta_{L}^{di}J_{dL(i)}j_{R}\right].
\label{eff-q}
\end{eqnarray}
($i$ runs over generations so that $q_i = u_i, d_i$ with
$u_i = u,c,t$ and $d_i = d,s,b$)
where the coefficients $\eta^q_k $ contain the $R_{p}\hspace{-1em}/\ \ $
SUSY parameters \cite{Kos-Kov}.
The color singlet currents $J_{q_{L/R}(i)}^{\mu}$ and $J_{d_{L/R}(i)}$
at the quark level are written as
$$
J_{q_{L/R}(i)}^{\mu } =\bar{q}_i \gamma^{\mu }P_{_{L/R}}q_{i},
\qquad \qquad
J_{d_{L/R}(i)}=\bar{d}_i P_{_{L/R}}d_{i}.
$$
where $P_{_{L,R}}=(1\mp \gamma_{5})/2$. The leptonic currents are written as
$$
j^{\mu }=\bar{e}\gamma^{\mu} P_{_{L}}\mu, \quad
\quad j_{_{L/R}}=\bar{e} P_{_{L/R}}\mu
$$

At the next step we rewrite Eq. (\ref{eff-q}),
specified at the quark level, in terms of the nucleon degrees of freedom.
This is usually achieved by utilizing the on-mass-shell
matching condition \cite{FKSS97} and gives
\begin{equation}
{\cal L}_{eff}^{N}=
\frac{G_F}{\sqrt{2}}\left[\bar{e}\gamma_{\mu }(1-\gamma _{5})\mu
\cdot J^{\mu }+
\bar{e}\mu \cdot J_{+}+\bar{e}\gamma_{5}\mu \cdot J_{-}\right].
\label{nucl1}
\end{equation}
where now the hadronic (nucleon) currents are
\begin{eqnarray}
J^{\mu } &=&\bar{N}\gamma ^{\mu }\left[ (\alpha _{V}^{(0)}+\alpha
_{V}^{(3)}\tau _{3})+(\alpha _{A}^{(0)}+\alpha _{A}^{(3)}\tau _{3})\gamma
_{5}\right] \,N,  \label{nucur} \\
J^{\pm } &=&\bar{N}\,\,\left[ (\alpha _{\pm S}^{(0)}+\alpha _{\pm
S}^{(3)}\tau _{3})+(\alpha _{\pm P}^{(0)}+\alpha _{\pm P}^{(3)}\tau
_{3})\gamma _{5}\right] \,N,  \nonumber
\end{eqnarray}
The coefficients $\alpha^{(0,3)}_K$,
with $K=S,V,A,P$, in Eq. (\ref{nucur}) include the nucleon form factors
(functions of the momentum transfer ${\bf q}^{2}$) for scalar, vector,
axial vector and pseudoscalar contributions, respectively (for their
definition see Ref. \cite{Kos-Kov}).
Since, however, the maximum momentum transfer ${\bf q}^{2}$ in
$\mu -e$ conversion is much smaller than the typical scale of nucleon
structure ($|{\bf q}| \approx m_\mu/c$, with $m_\mu=105.6 MeV$, the muon mass),
we can safely neglect the ${\bf q}^{2}$-dependence of the nucleon form factors.

\section{ Nucleon and nuclear structure dependence of the $\mu^--e^-$
conversion branching ratio}

One of the most interesting quantities in $\mu^--e^-$ conversion,
both from theoretical and experimental points of view, is the branching
ratio $R_{\mu e^-}$ defined in Eq. (\ref{def1}). The expression which gives $R_{\mu e^-}$ in the case of
the dominant coherent channel has been written as \cite{KV90}
        \begin{equation}
        R_{\mu e^-} = \rho\gamma,
        \label{III.2}
        \end{equation}
where $\rho$ is nearly
independent of nuclear physics~\cite{KFV97} and contains the
flavor-violating parameters mentioned before. Thus, e.g. for photon-exchange
$\rho$ is given by
\begin{equation}
\rho = (4\pi\alpha)^2\frac{\vert f_{M1}+f_{E0}\vert ^2+
\vert f_{E1}+f_{M0}\vert^2}{(G_Fm^2_{\mu})^2}
\label{rho}
\end{equation}
This expression contains the electromagnetic form factors $f_{E0}$,
$f_{E1}$,
$f_{M0}$, $f_{M1}$ parametrized in a specific elementary model~\cite{KFV97}.

The function $\gamma(A,Z)$ of Eq. (\ref{III.2}) includes about all the nuclear
information. By assuming that the total rate of the ordinary muon
capture rate is described by the Goulard-Primakoff
function $f_{GP}$, $\gamma(A,Z)$ is defined as
        \begin{equation}
        \gamma(A,Z)\equiv\gamma = \frac{E_e p_e}{m_\mu^2}
        \frac{ M^2}{G^2 Z f_{GP}(A,Z)},
        \label{III.3}
        \end{equation}
where $G^2 \approx 6$. Thus, the nuclear aspects of the $\mu^-\to e^-$
branching ratio $R_{\mu e^-}$ are mainly included in the matrix elements $M^2$
\cite{kos94}, which in the proton-neutron representation, are written as
        \begin{equation}
        M^2 = (M_p+ M_n)^2
        \label{ME2}
        \end{equation}

In general,
the transition matrix elements ${{\cal M}}_{p,n}$ in Eq. (\ref{ME2})
depend on the final nuclear state populated during the $\mu-e$ conversion.
For ground state to ground state transitions (${\it gs \to gs}$) in spherically
symmetric nuclei the following integral representation is valid
\begin{equation}
{\cal M}_{p,n} = 4\pi \int j_0(p_e r) \Phi_\mu (r) \rho_{p,n} (r) r^2  dr
\label{V.1}
\end{equation}
where $j_0(x)$ the zero order spherical Bessel function and
$\rho_{p,n}$ the proton (p), neutron (n) nuclear density normalized to
the atomic number $Z$ and neutron number $N$ of the participating
nucleus, respectively. The space dependent part of the muon
wave function $\Phi_{\mu}({\bf r})$, a spherically symmetric
function, can be obtained by solving numerically the Schr\"odinger
and Dirac equations with the Coulomb potential.

For the coherent rate in light nuclei the factorization approximation (see
Ref. \cite{KVF98}) is very good and $M^2_{\rm coh}$ can be expressed
in terms of the nuclear form factors $F_Z(q^2)$ (for protons) and $F_N(q^2)$
(for neutrons) which are easily estimated (for this channel only the ground
state wave function of the studied nucleus is required).
These form factors are defined as
        \begin{equation}\label{formZ}
        F_Z=\frac{1}{Z}\sum_j \hat j(j\Vert j_0(qr)\Vert j)
        \left(V_j^p\right)^2,
        \end{equation}
        \begin{equation}\label{formN}
        F_N=\frac{1}{N}\sum_j \hat j(j\Vert j_0(qr)\Vert j)
        \left(V_j^n\right)^2,
        \end{equation}
and contain the single-particle orbit occupancies $\left(V_j\right)^2$ for
the evaluation of which one must use a nuclear model in the proton-neutron
representation, e.g. QRPA (for $^{48}$Ti and $^{208}$Pb see Ref. \cite{KFV97}),
shell-model (for $^{27}$Al see Ref. \cite{SSK99}) etc.

By using the Lagrangian (\ref{nucl1}) one can deduce a similar expression
to that of Eq. (\ref{III.2}) for the $\mu-e$ conversion branching ratio as
\begin{equation}
R_{\mu e^-} = \frac{G^2_F } {2 \pi } \ \
\frac{ p_e E_e \ ({\cal M}_p + {\cal M}_n)^2}{ G^2 \ Z f_{GP}(A,Z)} \ \
{\cal Q}
\label{Rme}
\end{equation}
where
\begin{eqnarray}
{\cal Q} \,=\, 2|\alpha_V^{(0)}+\alpha_V^{(3)}\ \phi|^2 +
|\alpha_{+S}^{(0)}+\alpha_{+S}^{(3)}\ \phi|^2 + |\alpha_{-S}^{(0)}+
\alpha_{-S}^{(3)}\ \phi|^2
\nonumber  \\
+2\ {\rm Re}\{(\alpha_V^{(0)}+\alpha_V^{(3)}\ \phi)[\alpha_{+S}^{(0)}+\alpha_{-S}^{(0)}+
(\alpha_{+S}^{(3)}+\alpha_{-S}^{(3)})\ \phi] \}
\label{Rme.1}
\end{eqnarray}
with
\begin{equation}
\phi = ({\cal M}_p - {\cal M}_n)/({\cal M}_p + {\cal M}_n)
\label{phi}
\end{equation}
The quantity ${\cal Q}$ in Eq. (\ref{Rme.1}) depends weakly on the nuclear
parameters determining the factor $\phi$.
In fact, the terms depending on $\phi$ are small
and in practice the nuclear dependence of ${\cal Q}$ can be neglected.
Thus, the corresponding upper bounds on ${\cal Q}$ determine the
sensitivity to the \rp SUSY signals arriving at the detector in various
$\mu-e$ conversion experiments (see results below).

A usual approximate expression for the ratio $\phi$ of Eq. (\ref{phi}),
containing the nuclear parameters $A$, the atomic weight, and $Z$ the
total charge of the nucleus, is the following
\begin{equation}
\phi \approx {\tilde \phi} = (A-2Z)/A,
\label{phitil}
\end{equation}
The latter equation can be obtained assuming that
$M_p \approx Z \ F_Z \, \langle \Phi \rangle $ and
$M_n \approx N \ F_N \, \langle \Phi \rangle $
where $\langle \Phi \rangle $ represents the mean value of the muon wave function
$\Phi_{\mu}(r)$. For light and medium nuclei $F_Z \approx \ F_N$ and
Eq. (\ref{phitil}) is a good approximation.  For isoscalar nuclei, i.e.
$A=2 Z$, $\phi \approx {\tilde \phi} = 0$ (see Sect. 4).

\section{Results and Discussion.}

The pure nuclear physics calculations needed for the $\mu-e$ conversion studies
involve mainly the integrals of Eq. (\ref{V.1}). For the currently interesting nuclei
$Al$, $Ti$ and $Pb$ the results for $M_p$ and $M_n$ are shown in Table 1.
They have been calculated using proton densities $\rho_{p}$ from the electron
scattering data \cite{Vries} and neutron densities $\rho_{n}$ from the analysis
of pionic atom data \cite{Gibbs}. We have employed an analytical form for the muon wave
function obtained by solving the Schr\"odinger equation using the Coulomb potential
produced by the charge densities discussed before.
In this way the nucleon finite size was taken into consideration.
Moreover, we included vacuum polarization corrections as in Ref. \cite{Chiang}.

In solving the Schr\"odinger equation we have used modern neural
networks techniques \cite{Lagaris} which give the wave function $\Phi_\mu (r)$
in the analytic form of a sum over sigmoid functions. Thus, in Eq. (\ref{V.1})
only a simple numerical integration is finally required.
To estimate the influence of the non-relativistic
approximation on the muon wave function $\Phi_\mu({\bf r})$, we have also
determined it by solving the Dirac equation. The results do not significantly
differ from those of the Schr\"odinger picture.

In Table 1 we also show the muon binding energy $\epsilon_b$ (obtained
as output of the Dirac and Schr\"odinger solution) and the experimental values
for the total rate of the ordinary muon capture $\Gamma_{\mu c}$ taken from
Ref. \cite{Suzuki}.

Using values for $M_p$ and $M_n$ for a set of nuclei throughout
the periodic table \cite{Chiang} one can estimate the nuclear structure
dependence of the quantities $\rho$ and ${\cal Q}$. In Table 2
we show the results obtained for the ratio $\phi$ of Eq. (\ref{phi})
and its approximate expression $\tilde \phi$ of Eq. (\ref{phitil}).
We see that $\phi$ is very small ($-0.17 \le \phi \le 0.04$) and, because
the terms of the quantity $\cal Q$ depending on $\phi$ are also small,
we conclude that the nuclear dependence of $\cal Q$ can be
ignored as we have discussed in Sect. 2.

Since there is at maximum $\approx 15 \%$ difference between $\phi$ and
$\tilde \phi$ and because the isovector terms of the quantity $\cal Q$ are
small, especially for light nuclear systems, one can also use ${\tilde \phi}$ in
the expression (A,Z) dependence of $\cal Q$.

The results of Table 1 can be exploited for setting constraints on
the parameters of a specific gauge model predicting the $\mu-e$ process.
In Table 3 we quote the upper bounds for the quantities $\rho$ and
$\cal Q$ derived by using the recent experimental data on the
branching ratio $R_{\mu e^-}$ given in Eqs. (\ref{Rme-Ti}), (\ref{Rme-Pb})
and the expected experimental sensitivity of the Brookhaven experiment,
$R_{\mu e^-}<2\times 10^{-17}$ [see Eq. (\ref{Rme-Al})].
The limits of $\rho$ and ${\cal Q}$ quoted in Table 3
are improvements by about four orders of magnitude over the previous ones.

We should stress that the limits on the quantities $\rho$ of Eq. (\ref{rho})
and ${\cal Q}$ of Eqs. (\ref{Rme.1}), are the only constraints imposed by
the $\mu-e$ conversion on its underlying elementary particle physics.
One can extract upper limits on the individual lepton flavor violation
parameters (couplings of scalar, vector currents, neutrino masses etc.
\cite{Marci,kos94,KFV97,Kos-Kov}) under certain assumptions like the commonly
assumed dominance of only one component of the $\mu-e$ conversion
Lagrangian which is equivalent to constrain one
parameter or product of the parameters at a time. For example, in the
case of \rp  SUSY mechanisms, in order to learn about the size and
the regularities of possible violation of R-parity, one
needs some information on the parameters $\lambda, \lambda', \mu_i$ and
the sneutrino vacuum expectation values $\lg\tilde\nu_i\rg$
\cite{BFK}.
Using the upper limits for ${\cal Q}$ given in Table 3 we can derive
under the above assumptions the constraints on $\alpha^{(\tau)}_K$
of Eq. (\ref{Rme.1}) and the products of various \rp  parameters.
Thus, the bounds obtained for the scalar current couplings $\alpha^{(0)}_{\pm S}$
in the $R$-parity violating Lagrangian for the $^{27}$Al target
\cite{SSK99,SKSsub} are $|\alpha^{(0)}_{\pm S}| < 7 \times 10^{-10}$.
The limit for $\alpha^{(0)}_{\pm S}$ obtained with the data of Ti target \cite{Dohmen}
is $|\alpha^{(0)}_{\pm S}| < 1.1\times 10^{-7}$, i.e. more than two orders of
magnitude weaker than the limit of $^{27}$Al.

With these limits it is straightforward to derive constraints on the parameters
of the initial LFV Lagrangian. In Table 4 we list the most stringent
constraints on the products of the trilinear \rp couplings which we obtained
from the experimental limit on $\mu-e$ conversion in $^{48}Ti$ [see Eq.
(\ref{Rme-Ti})] and from the expected experimental sensitivity of MECO
detector using $^{27}$Al as a target material [see Eq. (\ref{Rme-Al})].
The corresponding constraints for $^{208}Pb$ are significantly weaker and
they are not presented in Table 4. In this table $B$ denotes a scaling factor
defined as $$ B = (R^{exp}_{\mu e}/7.0 \cdot 10^{-13})^{1/2},$$
which can be used for reconstructing the limits for the other
experimental upper limits on the branching ratio $R^{exp}_{\mu e}$.
As seen from Table 4, the $\mu-e$-conversion limits on the products
$\lambda\lambda'$ are significantly more stringent than those previously
known in the literature \cite{Rp-rev} and given in the 2nd column.
In Refs. \cite{Huitu,Kos-Kov} it was shown that except few cases,
the constraints on $\lambda' \lambda'$, $\lambda \lambda'$ and
$\lambda \lambda$ obtained from $\mu-e$ conversion data are better
than those derived from any other process.

As we have mentioned at the beginning, significantly better improvement on these
limits is expected from the ongoing experiments at PSI \cite{Dohmen} and even
better from the MECO experiment at Brookhaven \cite{Molzon}. This would make
the  $\mu-e$ conversion constraints better than those from the other
processes in all the cases.

Before closing we should note that the last four limits for $\lambda'\lambda$ in
Table 4 originate from the contribution of the strange nucleon sea. These
limits are comparable to the other $\mu-e$ constraints derived from the
valence quarks contributions.

\section{Summary and Conclusions.}

The transition matrix elements of the flavor violating $\mu^--e^-$
conversion are of notable importance in computing accurately the corresponding
rates for each accessible channel of this exotic process.
Such calculations provide useful nuclear-physics inputs for the expected
new data from the PSI and MECO experiments to put more severe bounds on the
muon-number-changing
parameters (isoscalar couplings, etc.) determining the effective currents
in various models that predict the exotic $\mu^--e^-$ process.

In the case of the R-parity violating interactions discussed here
we have studied all the possible tree-level contributions to the
$\mu-e$ conversion in nuclei
taking into account the nucleon and nuclear structure
effects. We found new important contribution to $\mu^- \to e^-$
originating from the strange quark sea in the nucleon which
is comparable with the usual contribution of the valence $u,d$ quarks.

From the existing data on $R_{\mu e^-}$ in $^{48}Ti$ and $^{208}Pb$
and the expected sensitivity of the designed MECO experiment \cite{Molzon}
we obtained stringent upper limits on the quantities $\rho$ and $\cal Q$
introduced in Eqs. (\ref{rho}) and (\ref{Rme.1}). They can be considered as
theoretical sensitivities of a particular
experiment to the $\mu^--e^-$ conversion
signals for various targets employed. Thus, these
quantities are helpful for comparing different $\mu-e$ conversion
experiments.
We also extracted the upper limits on the products of the trilinear \rp
parameters of the type $\lambda\lambda'$ which are significantly
more severe than those existing in the literature.
As to the other products, the following observation was formulated
in Refs. \cite{Huitu,Kos-Kov}. If the ongoing experiments
at PSI \cite{Dohmen} and Brookhaven \cite{Molzon}
will have reached the quoted sensitivities in the branching ratio
$R_{\mu e^-}$ then the $\mu-e$ constraints on all the products of
the \rp parameters $\lambda\lambda$, $\lambda'\lambda$,
$\lambda'\lambda'$ measurable in $\mu-e$ conversion
will become more stringent than those from any other processes.
This is especially important in view of the fact
that no comparable improvements of the other experiments
testing these couplings are expected in the near future.

\bigskip
\bigskip

\noindent
T.S.K would like to acknowledge partial support of this work
by Ioannina University grants and the NANP-99 Conference
Organizers for hospitality at Dubna.


%
\newpage
\begin{table}[ht]
\begin{center}
\begin{tabular}{rccccc}
\hline \hline
   &  &  &  &  &  \\
Nucleus & $|{\bf p}_e| \, (fm^{-1})$ & $\epsilon_b \, (MeV)$ &
$\Gamma_{\mu c} \, ( \times 10^{6} \, s^{-1})$ &
${\cal M}_p \, (fm^{-3/2})$ & ${\cal M}_n \, (fm^{-3/2})$  \\
\hline
   &  &  &  &  &  \\
$^{27}Al$  & 0.531 &  -0.470 &  0.71 & 0.047 & 0.045   \\
   &  &  &  &  &  \\
$^{48}Ti$  & 0.529 &  -1.264 &  2.60 & 0.104 & 0.127   \\
   &  &  &  &  &  \\
$^{208}Pb$ & 0.482 & -10.516 & 13.45 & 0.414 & 0.566   \\
\hline
\hline
\end{tabular}
\caption{Transition matrix elements (muon-nucleus overlap
integrals ${\cal M}_{p,n}$ of Eq. (\ref{V.1}))
evaluated by using the exact muon wave function obtained
via neural networks techniques. Other useful quantities
(see text) are also included.}
\end{center}
\end{table}

\vspace{0.5 cm}

\begin{table}
\begin{center}
\begin{tabular}{rrrr}
\hline
\hline
 & & &  \\
 A & Z & $\phi$(A,Z) &  ${\tilde \phi}$(A,Z) \\
\hline
 & & &  \\
 12. &  6.  &  0.000  &  0.000  \\
 24. & 12.  &  0.014  &  0.000  \\
 27. & 13.  &  0.000  & -0.037  \\
 32. & 16.  &  0.023  &  0.000  \\
 40. & 20.  &  0.037  &  0.000  \\
 44. & 20.  & -0.063  & -0.091  \\
 48. & 22.  & -0.083  & -0.083  \\
 63. & 29.  & -0.056  & -0.079  \\
 90. & 40.  & -0.054  & -0.111  \\
112. & 48.  & -0.108  & -0.143  \\
208. & 82.  & -0.152  & -0.212  \\
238. & 92.  & -0.175  & -0.227  \\
\hline
\hline
\end{tabular}
\caption{ The variation of the quantities $\phi$ [see Eq. (\ref{phi})]
and its approximate expression $\tilde \phi$ [see Eq. (\ref{phitil})]
through the periodic table. }
\end{center}
\end{table}

\vspace{0.5 cm}

\begin{table}
\begin{center}
\begin{tabular}{lccc}
\hline
\hline
 & & &  \\
Mechanism & \multicolumn{1}{c}{\large \bf $^{27}Al$}&
   \multicolumn{1}{c}{\large \bf $^{48}Ti$}&
   \multicolumn{1}{c}{\large \bf $^{208}Pb$} \\
\hline
 & & &  \\
 Photonic       &      $\rho$ $\le$ 4.6 $\times 10^{-18}$
                     & $\rho$ $\le$ 8.2 $\times 10^{-14}$
		     & $\rho$ $\le$ 3.2 $\times 10^{-12}$ \\
W-boson exchange&      $\rho$ $\le$ 5.8 $\times 10^{-19}$
                     & $\rho$ $\le$ 3.0 $\times 10^{-14}$
		     & $\rho$ $\le$ 1.1 $\times 10^{-12}$ \\
SUSY sleptons  &      $\rho$ $\le$ 1.8 $\times 10^{-18}$
                     & $\rho$ $\le$ 3.0 $\times 10^{-14}$
                     & $\rho$ $\le$ 1.1 $\times 10^{-12}$ \\
SUSY Z-exchange &      $\rho$ $\le$ 7.3 $\times 10^{-19}$
                     & $\rho$ $\le$ 0.7 $\times 10^{-14}$
                     & $\rho$ $\le$ 0.2 $\times 10^{-12}$ \\
\rp  SUSY   &        ${\cal Q}$ $\leq$ 5.10 $\cdot 10^{-19}$
                     & ${\cal Q}$ $\leq$ 1.25 $\cdot 10^{-14}$
                     & ${\cal Q}$ $\leq$ 2.27 $\cdot 10^{-13}$ \\
\hline
\hline
\end{tabular}
\caption{Upper limits on the elementary particle sector of the exotic
$\mu-e$ conversion branching ratio (quantity $\rho$ of Eq. (\ref{rho})
and quantity ${\cal Q}$ of Eq. (\ref{Rme.1}))
extracted by using the recent experimental data for the nuclear targets
$^{48}$Ti and $^{208}$Pb \protect\cite{Dohmen,Honec}
and the sensitivity of the MECO experiment for the $^{27}$Al target
\protect\cite{Molzon}. }
\end{center}
\end{table}

\begin{table}
\begin{center}
\begin{tabular}{cccc}
\hline
\hline
 &&& \\
Parameters &Previous bounds &  {\large \bf $^{48}Ti$}$\times B$ &
{\large \bf $ ^{27}Al$} $\times B$ \\
\hline
 &&& \\
$|\lambda '_{211}\, \lambda_{212}|$&$4.5\cdot 10^{-3}$   & $4.1\cdot 10^{-9}$& $5.5\cdot 10^{-11}$ \\
$|\lambda '_{311}\, \lambda_{312}|$&$6.0\cdot 10^{-3}$   & $4.1\cdot 10^{-9}$& $5.5\cdot 10^{-11}$ \\
$|\lambda '_{111}\, \lambda_{121}|$&$1.5\cdot 10^{-5}$   & $4.1\cdot 10^{-9}$& $5.5\cdot 10^{-11}$ \\
$|\lambda '_{311}\, \lambda_{321}|$&$6.0\cdot 10^{-3}$   & $4.1\cdot 10^{-9}$& $5.5\cdot 10^{-11}$ \\
$|\lambda '_{222}\, \lambda_{212}|$&$9.0\cdot 10^{-3}$   & $7.7\cdot 10^{-9}$& $1.0\cdot 10^{-10}$ \\
$|\lambda '_{322}\, \lambda_{312}|$&$1.2\cdot 10^{-2}$   & $7.7\cdot 10^{-9}$& $1.0\cdot 10^{-10}$ \\
$|\lambda '_{122}\, \lambda_{121}|$&$1.0\cdot 10^{-3}$   & $7.7\cdot 10^{-9}$& $1.0\cdot 10^{-10}$ \\
$|\lambda '_{322}\, \lambda_{321}|$&$1.2\cdot 10^{-2}$   & $7.7\cdot 10^{-9}$& $1.0\cdot 10^{-10}$ \\
\hline
\hline
\end{tabular}
\caption{The upper bounds on the products of the trilinear \rp
parameters $\lambda'_{ijk}\lambda_{lmn}$. Previous bounds in the 2nd column
are taken from Refs. \protect\cite{Huitu,Rp-rev}. The new $\mu^--e^-$
conversion bounds in the 3rd and the 4th columns are obtained using
the SINDRUM II data for ${}^{48}Ti$ target [see Eq. (\ref{Rme-Ti})] and
the expected sensitivity of the MECO target $^{27}$Al
[see Eq. (\ref{Rme-Al})]. These bounds are given for the scalar superpartner
masses $\tilde m = $ 100~GeV (the scaling factor $B$ is defined in the text). }
\end{center}
\end{table}
\end{document}